\documentstyle[12pt]{article}
\catcode`\@=11

\@addtoreset{equation}{section}

\def\section{\@startsection{section}{1}{\z@}{3.5ex plus 1ex minus
   .2ex}{2.3ex plus .2ex}{\large\bf}}

\newcommand{\be}{\begin{equation}}
\newcommand{\ee}{\end{equation}}
\newcommand{\bea}{\begin{eqnarray}}
\newcommand{\eea}{\end{eqnarray}}

\newcommand{\half}{{1\over2}}
\def\Idoubled#1{{\rm I\kern-.22em #1}}

\newcommand{\ID}{\Idoubled D}

\newcommand{\req}[1]{Eq.\ (\ref{#1})}

\begin{document}
\begin{titlepage}
%\vspace{.25in}
\begin{flushright}
%PACS 04.70.Dy\\
UNB Technical Report 96-04\\
%gr-qc/9410021\\
October 1996\\
\end{flushright}
\vspace{.25in}

\begin{center}
{\large\bf A Generalization of the Casson Invariant}\\
\vspace{15pt}
{\it by}\\
\vspace{5pt}
J.\ Gegenberg \\[5pt]
{\it Department of Mathematics and Statistics}\\
   {\it University of New Brunswick}\\
   {\it Fredericton, New Brunswick
   \it Canada E3B 5A3}\\
{[e-mail:  lenin@math.unb.ca]}\\
\end{center}
\vspace{20pt}
\begin{center}
{\bf Abstract}\\
\end{center}
A three dimensional supergravity theory which generalizes the super
IG theory of Witten and resembles the model discussed recently by
Mann and Papadopoulos is displayed.  The partition function is
computed, and is shown to be a three-manifold invariant generalizing
the Casson invariant.  

\end{titlepage} 

\section{Three
Dimensional Supergravity}

The theory of supergravity which I will consider is closely related to that 
constructed 
recently by Mann and Papadopoulos \cite{mannpap}.  This supergravity
model is a generalization of Witten's ``super IG" theory
\cite{witten1} and of the ``teleparallel theory" constructed by
Carlip and the present author \cite{cargeg}.

Although the supergravity theory can be written as supersymmetric Chern-Simons
theory, it will be displayed here in a more useful form, wherein it resembles 
(first-order) 3-d Einstein gravity coupled to families of bosonic and fermionic vector 
fields.

The bosonic field content consists of three 1-form fields $E^a, B^a,
C^a$ taking their values in the Lie algebra $so(3)$ and an SO(3)
connection 1-form $A^a$.  The fermionic fields are a pair of
spinorial 1-forms $\psi^i$, where the $so(3)$ indices $a,b,...=1,2,3$, and 
the $i,j,...=1,2$.  As we will see, the gravitational field is 
essentially determined by the $E^a$, while the the $\psi^i$ are gravitino 
fields.  I use the same conventions in as in \cite{mannpap}, namely that 
the gamma matrices $\gamma^a$ are pure imaginary so that 
\bea
\epsilon_{\alpha\gamma}\gamma_a{}^\alpha{}_\beta&=&\epsilon_{\alpha\beta}
\gamma_a{}^\alpha{}_\gamma,\\
\gamma^a\gamma^b&=&\delta^{ab}+i\epsilon^{abc}\gamma^c,\\
(\bar\psi)_\alpha&=& \psi^\beta\epsilon_{\alpha\beta},
\eea
where $\alpha,\beta,...=1,2$ are spinor indices.

The action for IISO(3) supergravity  
is 
\be
I=\int E^a\wedge F_a(A) + B^a\wedge D_AC_a 
+i\bar\psi^i\wedge\ID_A\psi^i
\ee
The operators $D_A,\ID_A$ are defined by
\bea
D_AB_a&:=&dB_a+\half\epsilon_{abc}A^b\wedge B^c,\\
\ID_A\psi^i&:=&d\psi^i-{ i\over4} A^a\gamma_a\wedge \psi^i.
\eea

The action is invariant under the supersymmetry transformations
\bea
\delta E^a&=&\half\bar\zeta^i\gamma^a\psi^i, \\
\delta\psi^i&=&\ID_A\zeta^i,
\eea
where the $\zeta^i$ are spinorial parameters.  As well, the action is invariant under
the IISO(3) gauge transformations:
\bea
\delta B^a=&D_A\rho^a+\half\epsilon^{abc}B^b\tau^c,\\
\delta C^a=&D_A\lambda^a+\half\epsilon^{abc}\tau^c,\\
\delta E^a=&D_A\xi^a+\half\epsilon^{abc}\left(E^b\tau^c+B^b\lambda^c+C^b\rho^c
\right),\\
\delta A^a=&D_A\tau^a.
\eea

The equations of motion of the theory are 
\bea
F^a(A):=dA^a+{1\over4}\epsilon^{abc}A_b\wedge A_c&=&0,\\
D_AB^a&=&0,\\
D_AC^a&=&0,\\
\ID_A\psi^i&=&0,\\
D_A E^a+\half\epsilon^{abc}B_b\wedge C_c-{1\over4}\bar\psi^i\wedge 
\gamma^a\psi^i&=&0.\label{DE}
\eea
The solutions of the equations of motion can be interpreted as a three-dimensional $N=2$ supergravity theory.  
The gravitino fields are the $\psi^i$, while the gravitational field consists 
of the spacetime triad $E^a$ and a compatible spin-connection $\omega^a$, 
determined by 
\be
dE^a+\half \epsilon^{abc}\omega_b\wedge E_c=0,\label{comp}
\ee
{\it i.e.} by 
\be
\epsilon^{abc}\left[\left(\omega_b-A_b\right)\wedge E_c-
B_b\wedge C_c\right]+\half\bar\psi^i 
\wedge\gamma^a\psi^i=0,
\ee
as can be seen from \req{DE} and \req{comp}.

\section{The Partition Function}

The partition function for the quantum super-IISO(3) theory is 
\be
Z=\int d\mu[\Phi]e^{-I[\Phi]},
\ee
where $d\mu[\Phi]$ is the measure in the configuration space of the fields 
$E,B,C,A,\psi$, denoted collectively by $\Phi$.
The partition function splits up:
\be
Z=\int d\mu[E,A]  e^{-\int E^a\wedge F_a(A)}Z[B,C]Z[\psi]
,\ee
where
\bea
Z[B,C]&:=&\int d\mu[B,C] e^{-\int B^a\wedge DC_a},\\
Z[\psi]&:=&\int d\mu[\psi] e^{-\int i\bar\psi^i\wedge\ID\psi^i}.
\eea
In the above, since the integral over $E$ has support only on 
regions of configuration space where $A$ is flat, the 
functional integrals $Z[B,C],Z[\psi]$ are evaluated at some flat $A$.   
\footnote{The calculation of the partition function was performed using the method 
outlined in \cite{part}, {\it i.e.} by parametrizing the fields $B,C,E$ and 
$\psi$ by their respective Hodge decompositions with respect to a flat SO(3) connection 
$\bar A$, and parametrizing $A$ itself by $A=\bar A+*D_{\bar A}*\alpha$, with 
$\alpha$ a closed so(3)-valued 2-form field.  The $*$ denotes the Hodge dual 
with respect to some Riemannian metric on the manifold $M$.  The  
functional integral measure of each field contains as a factor the inverse of 
the volume of the appropriate gauge orbit, which cancels the corresponding  
functional integral over the pure-gauge (exact) Hodge component.  The Jacobians 
and the Gaussian integrals over the bosonic fields give {\it absolute values} 
of powers of functional determinants; while `Gaussian integrals' over the 
fermionic fields have a specific sign \cite{birmingham}.}  
It turns out that the 
integral over $A$ cancels 
$Z[\psi]$ {\it up to a sign}, as in \cite{witten1}, while 
the $Z[B,C]=T(A)$, the Ray-Singer torsion for the flat connection $A
$.  Hence the partition function would be 
\be
Z=\sum_\alpha (-1)^{t_\alpha}T(A_\alpha)
\ee
where the index $\alpha$ is over the flat connections.  The exponents $t_\alpha
$ are $0$ or $1$.  

The partition function for pure bosonic IISO(3) gravity is 
\be
Z_b=\sum_\alpha \mid T(A_\alpha)\mid^2,
\ee
while for Witten's fermionic super-ISO(3) model, the partition function is
\be
Z_f=\sum_\alpha (-1)^{t^\alpha}.
\ee
Witten \cite{witten1} identifies the latter as the Casson invariant of the 
three-manifold.  The partition function $Z$ for super-IISO(3) theory bears 
rather the same relation to $Z_b$ and $Z_f$ as a complex number does to 
its modulus squared and its argument, respectively.  It is for this reason 
that the topological invariant $Z$ is interpreted as a new three-manifold 
invariant.

Work is currently in progress to understand the geometrical/topological 
meaning of $Z$.   

\bigskip\noindent
{\bf Acknowledgments}:  The author would like to thank George Papadopoulos 
for a useful discussion, and Steve Carlip for suggesting that a 
generalization of the Casson invariant might be constructed from quantum 
three-dimensional supergravity. 


\begin{thebibliography}{99}


\bibitem{mannpap}  R. Mann and G. Papadopoulos, ``Killing spinors, the 
adS black hole and I(ISO(2,1)) gravity", gr-qc/9606025 (1996).

\bibitem{witten1}  E. Witten, Nuc. Phys. B{\bf 323}, 113 (1989).

\bibitem{cargeg} S.\ Carlip and J.\ Gegenberg, Phys.\ Rev.\ 
{\bf D44}, 424 (1991). 

\bibitem{part} J. Gegenberg and G. Kunstatter, Ann. Phys. {\bf231}, 270 
(1994).

\bibitem{birmingham}  D. Birmingham, {\it et. al.}, Physics Reports {\bf 
209}, 129 (1991).


\end{thebibliography}
\end{document}